\documentclass[12pt]{iopart}
\usepackage{graphicx}

\begin{document}

\title[Cosmic Acceleration]{A Measurement of the Cosmic Expansion \\ Within our Lifetime}

\author{Fulvio Melia}

\address{Department of Physics, the Applied Math Program, and Department of Astronomy,
              The University of Arizona, Tucson, AZ 85721}
\ead{fmelia@email.arizona.edu}
\vspace{10pt}
\begin{indented}
\item[]July 16, 2021
\end{indented}

\begin{abstract}
The most exciting future observation in cosmology will feature a monitoring of the
cosmic expansion in real time, unlike anything that has ever been attempted before.
This campaign will uncover crucial physical properties of the various 
constituents in the Universe, and perhaps answer a simpler question
concerning whether or not the cosmic expansion is even accelerating. An unambiguous 
yes/no response to this query will significantly impact cosmology, of course, but 
also the standard model of particle physics. Here, we discuss---in a straightforward 
way---how to understand the so-called `redshift drift' sought by this campaign, and 
why its measurement will help us refine the standard-model parameters 
if the answer is `yes.' A `no' answer, on the other hand, could be more revolutionary,
in the sense that it might provide a resolution of several long-standing problems and 
inconsistencies in our current cosmological models. An outcome of zero redshift drift, 
for example, would obviate the need for a cosmological constant and render inflation 
completely redundant.
\end{abstract}
%
\vspace{2pc}
\par
%
\submitto{\EJP}
%
%
%

\section{Introduction}
Objects receding from us with the general Universal expansion become fainter with
time, and their spectra are redshifted according to their distance.  The rate at which
these quantities change is characterized by the expansion speed and acceleration, but
is scaled to the age of the Universe ($t_0\sim 13.5$ Gyr), which is considerably
longer than a human lifetime. It would therefore be farfetched to even consider
`watching the Universe expand in real time.'

And yet, there is increasing excitement and anticipation today at the prospect of actually
measuring the evolving redshift of distant sources via a campaign lasting several decades.
The idea of measuring the cosmic expansion in real time was first envisaged by Sandage 
(1962) over half a century ago, who estimated that galaxies would exhibit
spectroscopic velocity shifts (see Eq.~\ref{eq:Dv} below) of order $1$ cm s$^{-1}$
yr$^{-1}$. But he concluded that with optical and radio techniques available back then,
there would be no hope of measuring such relatively small changes, except over a very
long time (certainly longer than many decades). This type of cosmological experiment is
quite unique, however, because one does not need to pre-assume any particular model
to acquire the data and, more to the point, because the signal continues to grow larger
with time.

These are among the reasons why such a program has never been completely abandoned, and
today we find ourselves at the threshold of initiating several monitoring campaigns
expected to deliver compelling results by 2040. A principal motivation for this type
of observation is that it completely avoids the inherent difficulties in measuring
apparent magnitudes (Christiansen \& Siver 2012) which are subject to signal-to-noise
limitations and intrinsic source variability and evolution. The so-called `redshift drift'
expected from the Sandage Test is free of all such constraints.

Many therefore expect these monitoring campaigns to deliver an unprecedented improvement
in our ability to measure the cosmological parameters. There isn't much doubt about this,
but it is not widely recognized that there is an even simpler question to
be answered by these observations, whose outcome may resolve several inconsistencies in
our current models. Our goal in this paper is to present a straightforward explanation 
for the relevance of redshift drift to our comprehensive understanding in cosmology, 
and why a more precise measurement of the acceleration of the Universe 
may lead to a paradigm shift in our theoretical understanding by the time these monitoring 
campaigns release their initial results a few decades from now.

We shall begin with a pedagogical explanation of redshift drift and then convey the basic
idea of how its measurement may be used to refine our cosmological parameters in \S~2.
Beginning with \S\S~3 and 4, however, we shall focus more on the basic
question we are addressing in this paper, i.e., will the measured redshift drift 
confirm whether or not the Universe is actually accelerating. We shall discuss the 
significance of a zero redshift drift on our cosmological modeling in \S~5, and present 
our conclusions in \S~6.

\section{\label{sec:RedshiftDrift}Redshift Drift}
Cosmology today is based on the Friedmann-Lema\^itre-Robertson-Walker (FLRW) metric (see, 
e.g., Cook \& Burns 2009; Melia 2020) for a spatially homogeneous and isotropic 
three-dimensional space, expanding or contracting according to a time-dependent expansion 
factor, $a(t)$:
\begin{equation}
ds^2=c^2\,dt^2-a^2(t)\left[{dr^2\over (1-kr^2)}+
r^2(d\theta^2+\sin^2\theta\,d\phi^2)\right]\;.\label{eq:FLRW}
\end{equation}
\vskip 0.1in\noindent
This form of the metric is written using the coordinates of a comoving
observer, for whom $t$ is the cosmic time (and is the same everywhere),
$r$ is the comoving radius, which remains fixed for any source lacking
so-called peculiar motion, and $\theta$ and $\phi$ are the usual poloidal
and azimuthal angles. The geometric factor $k$, also known as the spatial
curvature constant, is $+1$ for a closed universe, $0$ for a flat universe,
and $-1$ for an open universe. All of the data today suggest that $k$ is very
likely zero (Planck Collaboration 2018), so we shall assume $k=0$ throughout this paper.
It is beyond the scope of this work to discuss the physical implication
of this measurement, but it is interesting to note that spatial curvature
is related to the local kinetic plus gravitational energy densities in the
Universe (Melia 2020), implying an important (and interesting!)
initial condition at the time of the Big Bang.

\begin{figure}[h]
\centerline{
\includegraphics[angle=0,scale=0.7]{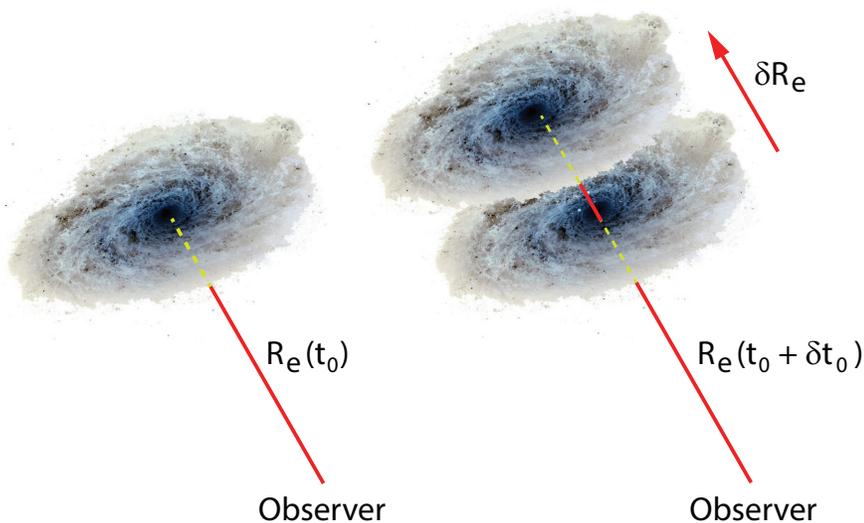}}
\vskip 0.2in
\caption{Schematic diagram showing the changing proper distance to a source, $R_{\rm e}$,
as the observer's time advances from $t_0$ to $t_0+\delta t_0$.}
\label{fig1}
\end{figure}

The coordinates $(ct,r,\theta,\phi)$ represent the perspective of a
{\it free-falling} observer. It is often also helpful to introduce the
{\it proper} coordinates that one may use to trace the physical location
of a source as a function of time. For example, the proper radius,
\begin{equation}
R(t)\equiv a(t)r\;,\label{Rt}
\end{equation}
is often used to express---not the co-moving (unchanging) distance $r$ between two
points but, instead---the increasing distance as the Universe expands, as illustrated
in Figure~\ref{fig1}.  This definition of $R$ is a consequence of Weyl's postulate (Weyl 1923),
which states that no two worldlines should ever cross (aside from any local peculiar
motion, if it exists).  According to Weyl, every physical distance in an FLRW cosmology
should thus be expressible as the product of a fixed comoving radius $r$, and the
aforementioned position-independent function of time, $a(t)$. One sometimes sees $R$
referred to as the areal radius---the radius of a two-sphere of symmetry---defined as
$R\equiv \sqrt{A/4\pi}$, independently of the coordinates, in terms of the area $A$
of the two-sphere (Nielsen \& Visser 2006; Abreu \* Visser 2010).

The most reliable information concerning the expansion factor $a(t)$ arrives to us in
the form of electromagnetic waves, shifted in frequency due to the combined influence of
kinematic and gravitationally induced redshift effects (Melia 2020). The null
geodesic equation describing the propagation of such waves along the $-\hat{r}$ direction,
with fixed $\theta$ and $\phi$, is easily obtained from Equation~(\ref{eq:FLRW}):
\begin{equation}
c\,dt=-a(t)\,dr\;.
\end{equation}
Thus, an electromagnetic signal emitted at $r_{\rm e}$, at time $t_{\rm e}$, will
reach the observer at time $t_0$ given by
\begin{equation}
\int_{t_{\rm e}}^{t_0}{dt\over a(t)}=r_{\rm e}\;.\label{eq:re}
\end{equation}
The comoving distance $r_{\rm e}$ is fixed in time, but the proper distance shown in
Figure~\ref{fig1} increases according to $R_{\rm e}(t_0)=a(t_0)r_{\rm e}$.

Information concerning the redshift of light emitted by the source is encoded in
Equation~(\ref{eq:re}), which tells us how $t_0$ changes as a function of $t_{\rm e}$
due to the evolution of $a(t)$ between these two times. For example, if we consider the
emission and detection of two crests of the wave, one at $t_{\rm e}$ and $t_0$, and the
second at $t_{\rm e}+\delta t_{\rm e}$ and $t_0+\delta t_0$, then we see from
Equation~(\ref{eq:re}) that
\begin{equation}
\int_{t_{\rm e}+\delta t_{\rm e}}^{t_0+\delta t_0}{dt\over a(t)}=
\int_{t_{\rm e}}^{t_0}{dt\over a(t)}\;.
\end{equation}
And given that $a(t)$ changes only by a miniscule amount during one period of a typical
wave, we may carry out these integrations trivially and find that
\begin{equation}
{\delta t_{\rm e}\over a(t_{\rm e})}={\delta t_0\over a(t_0)}\;.
\end{equation}
The conventional definition of {\it redshift}, $z$, is based on the fractional
increase in wavelength,
\begin{equation}
z\equiv {\lambda_0-\lambda_{\rm e}\over \lambda_{\rm e}}\;.
\end{equation}
Therefore, since the frequency of the wave is $\nu=c/\lambda$, and
\begin{equation}
{\nu_{\rm e}\over\nu_0}={\delta t_0\over \delta t_{\rm e}}\;,
\end{equation}
we arrive at one of the most fundamental equations in cosmology (Weinberg 1972):
\begin{equation}
1+z={a(t_0)\over a(t_{\rm e})}\;.\label{eq:z}
\end{equation}

\begin{figure}[h]
\centerline{
\includegraphics[angle=0,scale=1.0]{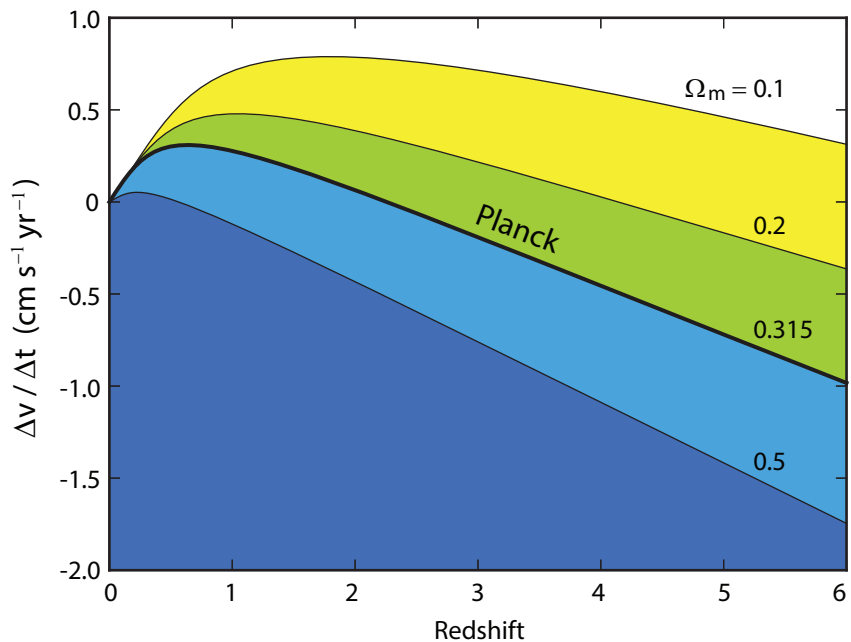}}
\vskip 0.1in
\caption{Spectroscopic velocity shift $\Delta v/\Delta t$ associated with the
redshift drift predicted by the {\it Planck}-$\Lambda$CDM model ($k=0$,
$\Omega_{\rm m}=0.315$, $H_0=67.4$ km s$^{-1}$ Mpc$^{-1}$; thick black line),
and several variations with alternative values of $\Omega_{\rm m}$ (indicated in
the plot). In every case, dark energy is assumed to be a cosmological constant, $\Lambda$,
with $w_{\rm de}=-1$ and $\Omega_\Lambda=1-\Omega_{\rm m}$.}
\label{fig2}
\end{figure}

Of course, this relation gives us the redshift corresponding to cosmic evolution over
millions and billions of years. It is hardly useful as a probe of the change occurring
over a mere human lifetime. It is necessary for us to derive from Equation~(\ref{eq:z})
an expression yielding the incremental changes in $z$ expected during a much shorter
time interval $\delta t_0$. Differentiating Equation~(\ref{eq:z}) with respect to
the observer's time $t_0$, we find that
\begin{equation}
{dz\over dt_0}=[1+z(t_0)]H(t_0)-{a(t_0)\over a(t_{\rm e})^2}{da(t_{\rm e})\over dt_{\rm e}}
{dt_{\rm e}\over dt_0}
\end{equation}
where, by definition, the Hubble parameter is given as
\begin{equation}
H(t)={1\over a(t)}{da(t)\over dt}\;.\label{eq:Hubble}
\end{equation}
And since
\begin{equation}
dt_0=[1+z(t_0)]\,dt_{\rm e}\;,
\end{equation}
we arrive at an equally elegant and simple expression for the {\it redshift drift}
expected over an observable interval of time:
\begin{equation}
{dz\over dt_0}=(1+z)H_0-H(z)\;.\label{eq:dzdt0}
\end{equation}
To simplify the notation somewhat, we have used the definitions $z\equiv z(t_0)$ and
$H_0\equiv H(t_0)$.

During a monitoring campaign, the surveys will measure the spectroscopic velocity
shift, $\Delta v$, defined in terms of the redshift drift $\Delta z$ over
an observation time $\Delta t$:
\begin{equation}
\Delta v\equiv {c\Delta z\over 1+z}={c\Delta t\over 1+z}\,{dz\over dt_0}\;.\label{eq:Dv}
\end{equation}
The expectation within the cosmology community in regard to the measurement 
of the redshift drift (i.e., $\Delta z$ over an interval of time $\Delta t$) is that
this equation may be used together with the new data to optimize the standard model's 
fit to the velocity shift as a function of redshift, thereby `measuring' the cosmological 
parameters to very high precision. The model dependence in Equation~(\ref{eq:Dv}) enters 
exclusively through the Hubble parameter $H(z)$ in Equation~(\ref{eq:dzdt0}) which, for 
the standard model ($\Lambda$CDM), is written as follows:
\begin{equation}
H(z)\equiv H_0E(z)\;,\label{eq:Hz}
\end{equation}
where
\begin{equation}
E^2(z)\equiv \Omega_{\rm m}(1+z)^3+\Omega_{\rm r}(1+z)^4+\Omega_{\rm de}
(1+z)^{3(1+w_{\rm de})}\;.\label{eq:E}
\end{equation}
This expression uses the standard ratios $\Omega_i\equiv \rho_i/\rho_{\rm c}$ of the
energy density for species $i={\rm m}$ (matter), $i={\rm r}$ (radiation) and
$i={\rm de}$ (dark energy), in terms of the current critical density $\rho_{\rm c}\equiv 
3c^2H_0^2/8\pi G$, and the dark-energy equation-of-state parameter, $w_{\rm de}\equiv 
p_{\rm de}/\rho_{\rm de}$, with $p_{\rm de}$ the pressure.

For example, in the simplified approach of assuming a spatially flat Universe (i.e.,
$k=0$) and dark energy in the form of a cosmological constant (Melia 2020) $\Lambda$
(with $w_{\rm de}=-1$), the monitoring of $\Delta v$ should provide a direct measurement of
$\Omega_{\rm m}$, and therefore of $\Omega_{\rm de}\equiv\Omega_\Lambda=1-\Omega_{\rm m}$.
To illustrate the potential for carrying out this groundbreaking work, we show in Figure~\ref{fig2}
the variation of $\Delta v/\Delta t$ (in units of cm s$^{-1}$ yr$^{-1}$) with redshift and
the matter density $\Omega_{\rm m}$. The version of $\Lambda$CDM with the parameters optimized
by {\it Planck} (Planck Collaboration 2018) (conventionally referred to as 
{\it Planck}-$\Lambda$CDM) is
highlighted on this plot with a thick black curve. Notice, e.g., that the redshift drift with
time is positive at low redshifts, and then turns negative for the more distant sources. This
unambiguous prediction by the standard model is simply based on the temporal evolution of
the matter ($\rho_{\rm m}$) and dark-energy ($\rho_{\rm de}$) densities, which sees the Universe
dominated by the former at $z > 0.7$, giving way to the latter towards the present. In
$\Lambda$CDM, dark energy functions as an agent of acceleration, whereas a matter-dominated
cosmos is always decelerating.

Measuring a spectroscopic velocity shift of $< 1$ cm s$^{-1}$ yr$^{-1}$ sounds very
challenging, but as we shall discuss in \S~4, the idea of observing redshift drift has come
a long way since the initial proposal by Sandage (1962) six decades ago. Many
groups have made projections for the next generation of telescope facilities, and have
concluded that a few thousand hours of observing time spread over a baseline of several
decades should provide a very compelling signal. Indeed, as we shall see shortly, the most
optimistic predictions show that the European Extremely Large Telescope (ELT) (Liske et al. 2014)
(but also the Square Kilometer Array, SKA; Kloeckner et al. 2015) will provide a $3\sigma$
result in under two decades. Before we consider the details of this project, 
however, let us first direct our attention to what may be the simplest question to be 
answered by a measurement of the redshift drift.

\section{Is the Universe Accelerating?}
Recent trends have started to indicate the possibility that a simple yes/no 
answer to this question may have an impact on several areas in cosmology and the standard 
model of particle physics, so let us now focus on this particular issue.

\begin{figure}
\centerline{
\includegraphics[angle=0,scale=1.0]{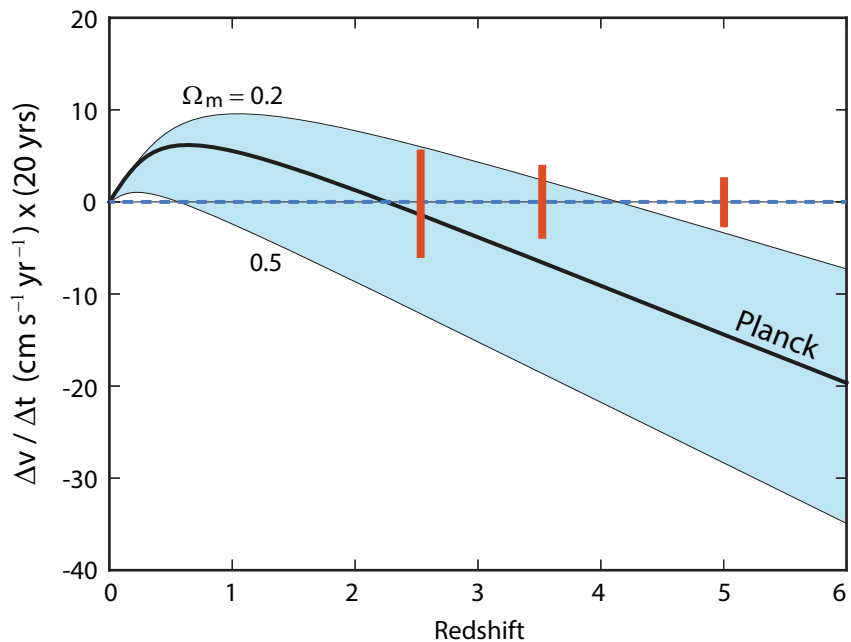}}
\vskip 0.1in
\caption{Spectroscopic velocity shift $\Delta v/\Delta t$ associated with three of the
$\Lambda$CDM models shown in Figure~\ref{fig2}, i.e., {\it Planck}-$\Lambda$CDM,
$\Omega_{\rm m}=0.2$
and $\Omega_{\rm m}=0.5$, compared with an FLRW Universe expanding at a constant rate
(blue long-dashed line), which would exhibit strictly zero redshift drift at all redshifts.
The velocity shifts are calculated over a twenty-year baseline, consistent with the expected
errors (red bars) after 20 years of monitoring with the ELT-HIRES (Liske et al. 2014).}
\label{fig3}
\end{figure}

The conventional wisdom developed over the past three decades would suggest that 
the Universe is currently accelerating.  Certainly, the analysis of Type Ia supernovae 
(Perlmutter et al.  1998; Riess et al. 1998; Schmidt et al. 1998) has provided a convincing, 
if not unchallenged (Melia 2020), argument that the Universe has undergone several phases of
deceleration and acceleration consistent with Equations~(\ref{eq:Hz}) and (\ref{eq:E}). But
the growing tension between the standard model's predictions and our ever improving observations
is beginning to cast a big shadow over its formerly perceived infallibility. For example, the
Hubble constant, $H_0$, characterizing the expansion rate and cosmic distance scale (Cadmus 1999),
cannot find a self-consistent value from the combined measurement of the cosmic microwave
background (CMB) (Planck Collaboration 2018), which gives $67.4\pm0.5$ km $\rm s^{-1}$ $\rm Mpc^{-1}$,
and local Type Ia supernovae calibrated with the Cepheid distance ladder (Riess et al. 2019),
which instead yield $74.03\pm1.42$ km $\rm s^{-1}$ $\rm Mpc^{-1}$. Part of the
problem is that this $4$--$5\sigma$ disparity seems to get worse as the sensitivity of the 
instruments gets better.

But there are many indications, besides the problem with the Hubble constant,
that the observations may not be fully consistent with an accelerating Universe. Indeed, 
the standard model suffers from several seemingly insoluble hurdles and inconsistencies at 
a very fundamental level, in spite of the general success it has enjoyed accounting for most 
of the observations over an extended period of time. We shall return to this shortly, after 
first describing how the upcoming measurement of the redshift drift will hopefully bring this 
decades-long debate to a close.

The distinction between an FLRW Universe expanding at a constant rate and one undergoing
some form of acceleration or deceleration is quite simple to understand (Melia 2016).
If $a(t)\propto t$, meaning no acceleration, Equation~(\ref{eq:Hubble}) gives $H(t)=1/t$,
while Equation~(\ref{eq:z}) shows that $1+z=t_0/t_{\rm e}$. Thus,
\begin{equation}
H(t_{\rm e})=H(t_0)[1+z(t_0)]\;,\label{eq:Hte}
\end{equation}
and
\begin{equation}
{dz\over dt_0}=0\;,
\end{equation}
so that $\Delta v=0$. In other words, a Universe without any acceleration has strictly
zero redshift drift at all redshifts. A source with a measured redshift at the time of its
discovery will retain that same redshift forever in a Universe expanding at a constant
rate. This is indicated by the blue long-dashed line in Figure~\ref{fig3}, along with
the anticipated measurement precision (red bars; see \S~4 below) and a comparison of this
outcome with three of the curves in Figure~\ref{fig2}.

\section{The Next Generation of Telescopes}
Today's most powerful telescopes can attain a precision in measuring redshifts better than
$\sim 10^{-7}$, as used in searching for planets or measuring time-varying fundamental
constants (Lovis \& Pepe 2007). Typical large-redshift surveys, such as the Sloan Digital
Sky Survey at Apache Point Observatory, do worse than this, which means that the best one
can hope for in measuring $\Delta v$ is no better than a few meters s$^{-1}$---nowhere
near the level of precision required in Figure~\ref{fig3}. Instead of discerning between
an accelerating and non-accelerating Universe in $\sim 20$ years, one would need to wait
many centuries to even begin seeing the cosmic expansion in `real' time.

\begin{figure}
\centerline{
\includegraphics[angle=0,scale=0.28]{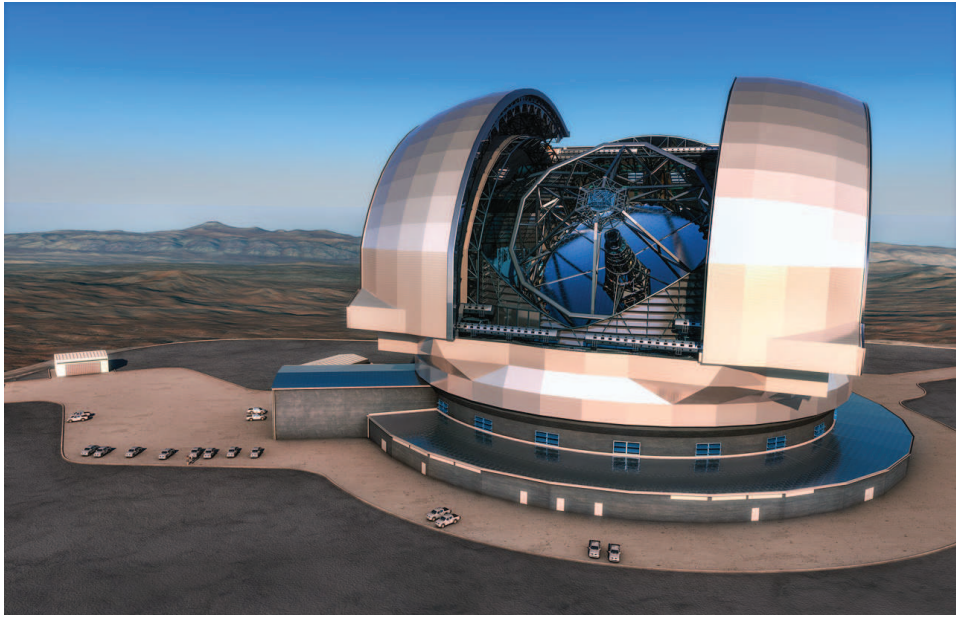}}
\vskip 0.1in
\caption{Artist's impression of the European Extremely Large Telescope, a 39-meter aperture
optical and infrared facility under construction on Cerro Armazones in the Chilean Atacama Desert.
(Courtesy ESO/L. Cal\c{c}ada)}
\label{fig4}
\end{figure}

This situation is changing dramatically with the development of new telescopes, however,
such as the ELT, with an order of magnitude
improvement in light-gathering power (Fig.~\ref{fig4}), advances in adaptive optics to mitigate
atmospheric distortions, and new technologies in quantum optics to facilitate the imprinting of
ultra-stable spectra. Such devices are expected to measure individual redshifts with
an order of magnitude improvement in the precision available today. With the enhanced
statistics from the many thousands of new, faint quasars that will be detected along the way,
their overall capability to measure $\Delta v$ should reach $\sim 10$ cm s$^{-1}$, right in the
ball park of Figure~\ref{fig3}. Integrating over a baseline of one to two decades should
therefore yield a signal-to-noise ratio sufficient to disentangle the various profiles shown
on this plot---in {\it less than a human lifetime}.

According to Liske et al. (2008), the High-Resolution Spectrograph (HIRES) on the
ELT is expected to observe a spectroscopic velocity shift, $\Delta v$, with an uncertainty of
\begin{equation}
\sigma_{\Delta v} = 1.35\, {2370\over {\rm S/N}}\sqrt{{30\over N_{\rm QSO}}}\left({5
\over 1+z_{\rm QSO}}\right)^\alpha\;{\rm cm}\;{\rm s}^{-1}\;,
\end{equation}
where $\alpha=1.7$ for $z\le 4$, and $\alpha=0.9$ for $z>4$, with ${\rm S/N}\approx 1,500$
after 5 years of monitoring $N_{\rm QSO}=10$ quasars in each of three redshift bins at
$z=2.5$, $3.5$ and $5.0$. With a baseline of 20 years, these uncertainties (shown as red
bars in Fig.~\ref{fig3}) are approximately $6$, $4$ and $3$ cm s$^{-1}$, respectively. Thus,
if the Universe is expanding at a constant rate (blue long-dashed line), we should see a
$\sim 3\sigma$ difference between a measurement of zero and the velocity shift
predicted by {\it Planck}-$\Lambda$CDM over a span of just two decades. The difference
continues to increase with integration time beyond this limit.

\section{The Impact of a Measured Acceleration}
The standard cosmological model we use to interpret the data today has grown
and evolved over several decades, though always based firmly on the FLRW metric shown in
Equation~(\ref{eq:FLRW}). No other attempt at solving Einstein's field equations in general
relativity has come close to supplanting this metric as the most likely description of the
cosmic spacetime. 

But look carefully again at the coefficients in this metric, notably the significant impact
of the expansion factor $a(t)$ on yielding changes in time as a function of position. Whereas
Einstein's equations provide us with this formal expression, they do not tell us much about
$a(t)$ itself, unless we introduce specific information about the source of gravity, in the 
form of a stress-energy tensor. And there is no fundamental theory we can rely on to do
this unambiguously. 

Much of the effort in observational cosmology since the 1960's has thus been directed at
identifying the various constituents that make up the fabric of the Universe---the stars,
galaxies, and gas between them, the background radiation and whatever else is suggested
by each new generation of higher precision instrumentation. As noted earlier, today we
believe the `cosmic fluid' contains baryonic matter, an unidentified dark matter, radiation,
and a very mysterious `dark energy.' But we cannot escape the fact that all of this
description, which is essential in determining the behaviour of $a(t)$, is largely empirical.
And this limitation continues to thwart our attempts at producing a complete theoretical
understanding of the origin and evolution of the Universe.

We can see this via Equation~(\ref{eq:E}), for example, which characterizes the Hubble
parameter---and thus the expansion rate of the Universe---in terms of several unknown
factors, such as the density ratios of the various constituents and the equation-of-state
of dark energy. Our cosmological measurements help us optimize their values, but how
do we make sense of $\Omega_{\rm de}$ being, say, $0.69$ or $0.72$? How does this 
differentiation inform the underlying theory? To this point, everything we know about 
the evolution of the Universe is based on measurements of integrated distances using 
standard candles, or the assumed size of large-scale structures, such as galaxy clusters, 
or the modeled growth of temperature variations in the CMB. But We're not even sure that 
this empirical approach makes sense.

What we mean by this is that a completely unavoidable prediction of the formulation in
Equation~(\ref{eq:E}) is that the Universe must undergo various phases of accelerated and
decelerated expansion. The reason for this is quite simple to understand. The various
constituents represented by $\Omega_{\rm m}$, $\Omega_{\rm r}$ and $\Omega_{\rm de}$
evolve over time at different rates from each other. For example, if dark energy were 
a cosmological constant, the quantity $\Omega_{\rm de}(1+z)^{3(1+w_{\rm de})}=\Omega_{\rm de}$ 
would never change, while both $\Omega_{\rm m}(1+z)^3$ and $\Omega_{\rm r}(z+z)^4$ decrease as 
the Universe ages (i.e., as $z$ decreases). And since their respective contributions
to the equation-of-state in the cosmic fluid must therefore also evolve with time, the
expansion rate of the Universe cannot itself remain constant, which therefore also means
that $dz/dt_0$ in Equation~(\ref{eq:dzdt0}) cannot be zero. But we have never confirmed
this fundamental property of the standard model.

How wonderful it would be to finally prove this feature with the actual measurement of
a redshift drift! At the very least, we would then be certain that a formulation like
that expressed in Equation~(\ref{eq:E}) does indeed make sense, and that our current
inability to relate the measured values of the unknown parameters to some overarching
theory will eventually give way to a much deeper understanding. In concert with such
a confirmation, we could then set about actually measuring quantities such as $\Omega_{\rm de}$
via Equations~(\ref{eq:dzdt0}), (\ref{eq:Hz}) and (\ref{eq:E}) with the ever improving
redshift-drift signal over time.

As thrilling as such an outcome would be, an argument can be made that the result would be
even more surprising---and very perplexing---if it turns out that $dz/dt_0=0$. That would 
be completely incomprehensible in the context of Equation~(\ref{eq:E}). It would imply
that such a formulation of the Hubble parameter is, at best, merely an approximation,
and that in reality $H(z)$ is given by the much simpler expression in Equation~(\ref{eq:Hte}),
which has only one unknown factor, i.e., the value of the Hubble constant today.

This possibility has been explored for over a decade now, largely focusing
on the mitigation of several growing tensions between the standard model and other types
of observation, as we shall describe shortly. It turns out that a constant expansion rate may
actually fit the data better than the variable one predicted by Equation~(\ref{eq:E})
(see, e.g., Melia 2020), so there are good reasons to believe that a measurement of 
zero redshift drift is not entirely out of the question.

This is what we have been highlighting throughout this paper, i.e., that an
important question to be answered by the redshift-drift measurements is not just which of 
the curves in Figure~\ref{fig2} best represents the data, but simply whether the drift is 
different from zero (Fig.~\ref{fig3}). The answer to this question is a straightforward 
yes or no. And zero redshift drift could alleviate several otherwise insurmountable problems 
with the standard model.

To begin with, the interplay between $\Omega_{\rm m}$ and $\Omega_{\rm de}$ in Equation~(\ref{eq:E}) 
that we have been discussing produces several alternating phases of deceleration and acceleration 
throughout cosmic history, but with the highly unlikely coincidence that we still end up with 
$H_0=1/t_0$ today, within the measurement errors (Melia 2020). To achieve this endpoint within 
the context of {\it Planck}-$\Lambda$CDM, dark energy needs to be a cosmological constant, 
with $w_{\rm de}=-1$ and $\Omega_{\rm de}=\Omega_\Lambda\approx 0.7$. As is well known, however, 
the required value of $\rho_{\rm de}=\rho_\Lambda$ is strikingly inconsistent with the much 
larger zero-point energy suggested by quantum field theory.  Depending on how one handles all 
the energy contributions, the disagreement can be as high as 120 orders of magnitude, a situation 
sometimes described as `the largest discrepancy between theory and experiment in all of science' 
(Adler et al. 1995; Harvey \& Schucking 2000).

In a constantly expanding cosmos, on the other hand, dark energy could not be a cosmological
constant. It would be a dynamic entity, akin to other particles and fields present in the
cosmic fluid. There is even a hint that some of it may have decayed in the early Universe to
produce the baryonic matter we see today (Melia 2020). Regardless, in the event that
the measured redshift drift turns out to be zero, dark energy's equation-of-state (different from
$p_{\rm de}=-\rho_{\rm de}$) and its coupling to other known particles, would provide significant
justification for its inclusion in an extended model of particle physics.

Dark energy impacts the cosmic dynamics towards the end of the Universe's timeline, as seen from
our perspective today. But there is an equally troubling and persistent issue plaguing the
standard model in the early Universe, typically referred to as the `horizon' problem. In fact,
there are several such problems. The first is associated with the temperature ($T_{\rm cmb}$)
of the CMB, which appears to be uniform across the sky (Planck Collaboration 2018), 
save for tiny fluctuations on the order of 1 part in $10^5$. The standard model, with 
the Hubble parameter in Equation~(\ref{eq:Hz}), would have experienced very rapid 
deceleration prior to the last scattering surface at $z_{\rm cmb}\sim 1080$, preventing 
the Universe from expanding sufficiently by that time for us to see this level of 
uniformity in $T_{\rm cmb}$ today.

A very brief period of inflation prior to $z_{\rm cmb}$ was proposed as a means of providing the
missing expansion (Guth 1981). But in the intervening four decades, attempts at identifying the
inflaton field responsible for this effect still have not determined its potential, and we cannot
claim to understand exactly how inflation works---or whether it is even consistent with all of the
observations. A careful analysis of the latest {\it Planck} data release (Melia \&
L\'opez-Corredoira 2018) shows that several-large angle anomalies associated with the 
aforementioned temperature fluctuations are probably due to a cutoff in the primordial 
power spectrum. But these anisotropies are attributed to quantum fluctuations in the 
inflaton field, and such a cutoff signals the time at which inflation could have started. 
Unfortunately, the data suggest that this initiation would have occurred too far beyond 
the Big Bang to have allowed inflation to simultaneously solve the temperature horizon 
problem and produce the fluctuations observed in the CMB (Liu \& Melia 2020).

A second horizon problem, emerging with the recent discovery (Aad et al. 2012) of the 
Standard Model Higgs boson, may be even worse. According to standard particle physics, 
the Universe may have undergone several phase transitions, in addition to the inflationary 
event we have just described, presumably driven by the separation of the strong and 
electroweak forces in grand unified theories. Now that we know the Higgs particle exists, 
it is very likely that the Universe underwent another (electroweak) phase transition at 
$T=159.5\pm1.5$ GeV. In {\it Planck}-$\Lambda$CDM, this would have occurred 
$\sim 10^{-11}$ seconds after the Big Bang. During this event, the electric and weak 
forces separated, and the fermions gained mass (Englert \& Brout 1964; Higgs 1964). The latter,
however, depends on the vacuum expectation value of the Higgs field. But this too appears 
to be uniform across the sky, since we do not see any evidence of fermion masses varying 
from one location to another. And regardless of whether or not inflation happened, the 
Universe would have resumed its deceleration afterwards, creating another horizon problem 
at $t\sim 10^{-11}$ seconds---this time with respect to the Higgs vacuum expectation value. 
This problem has been known since Zel'dovic et al. (1975) and 
Kibble (Kibble 1976) suggested the possibility that domain walls might have been 
created as a result of such scale transitions across the Universe, constituting topological 
defects with significant observational consequences (Vilenkin \& Shellard 1994;
Lazanu et al. 2015; Sousa \& Avelino 2015) in $T_{\rm cmb}$.  But no evidence of such 
artifacts has ever been found, nor has any viable explanation been offered within the 
context of the standard model to account for the uniformity of the Higgs field vacuum 
expectation value across the Universe.

But all of these horizon problems will disappear should 
the search for redshift drift produce a null result. These problems emerge only in cosmological
models, such as {\it Planck}-$\Lambda$CDM, that predict an early period of decelerated expansion.
But if the Universe has been expanding at a constant rate, it would always have had sufficient
time to expose proper distances beyond the most distant sources we see today. And this feature
is independent of the redshift of these sources or structures, including the last-scattering surface
at $z_{\rm cmb}$, and the apparent horizon at the time of the electroweak phase
transition (Melia 2020).

Between these two limits---where dark energy and $H_0$ appear to be problematic in the local
Universe, and where cosmic deceleration would necessarily have led to a series of horizon
inconsistencies at high redshifts---the standard model also suffers from a long-standing
time-compression problem associated with the premature formation of structure. Supermassive
black holes are seen (Banados et al. 2018) at redshifts exceeding $\sim 7.5$ which, in
{\it Planck}-$\Lambda$CDM corresponds to about 690 Myr after the Big Bang, barely a few
hundred Myr after the formation of Pop III and II stars. Conventional astrophysics, however,
would indicate that a $10\;M_\odot$ seed, created by a Pop II/III supernova, should have taken
at least 820 Myr to grow via Eddington-limited accretion (Melia 2020). The timeline thus
appears to have been overly compressed by a factor $\sim 2$ in the standard model, at least beyond
$z\sim 6$.  Attempts at addressing this issue invariably invoke exotic `fixes,' such as imposing
anomalously high accretion rates (Volonteri \& Riess 2005; Inayoshi et al. 2016), or the 
creation of very massive ($\sim 10^5\;M_\odot$) seeds (Alexander et al. 2014). But no direct 
evidence of such unconventional processes and events has ever been seen. A similar problem 
has also been observed via the premature formation of galaxies, some of which appeared 
beyond $z\sim 10-12$, much earlier than expected within the standard 
model (Steinhardt et al. 2016).

Contrasting with these inconsistencies, the formation of high-redshift quasars and galaxies
fits extremely well within the timeline associated with an FLRW Universe expanding at a constant
rate. In this case, a redshift of $\sim 15$---the start of the so-called Epoch of Reionization,
when Pop III and II stars began replenishing the interstellar medium with ionizing radiation---
corresponds to an age of $\sim 878$ Myr, as opposed to $\sim 400$ Myr in the standard model.
Most importantly, a $10\;M_\odot$ seed created at that time would have easily grown into a
$\sim 10^9\;M_\odot$ black hole at $z=7.5$ (i.e., $t\sim 1.65$ Gyr) via conventional
Eddington-limited accretion (Melia 2020). A measurement of zero redshift drift
could thus also cure the time-compression problem in {\it Planck}-$\Lambda$CDM.

These principal illustrations already highlight what is at stake with a precise
measurement of the cosmic acceleration, but the derived benefits extend beyond this brief 
survey. For example, if it turns out that the expansion rate is constant, then
(i) the concept of cosmic entropy would be simplified, eliminating the so-called initial entropy 
problem, in which one must scramble to explain how and why the entropy in the standard model 
began with an anomalously low value and increased rapidly by the time the CMB was produced 
(Melia 2021a); (ii) it would also help to explain how quantum fluctuations in an incipient 
scalar field could have classicalized and turned into density fluctuations growing under their 
own self-gravity (Melia 2021b); and (iii) it would avoid the so-called 
trans-Planckian anomaly (Martin \& Brandenberger 2001; Brandenberger \& Martin 2013), 
in which quantum fluctuations would otherwise have mysteriously emerged out of the 
Planck domain into the semi-classical Universe without an explanation from quantum mechanics 
and without a viable theory of quantum gravity.

\section{Conclusion}
{\it Planck}-$\Lambda$CDM has been quite successful in accounting for a broad range of
cosmological observations, but careful scrutiny reveals several major fundamental
problems with its theoretical foundation. Some of these issues have been with us for over
half a century. Moreover, as the precision of our instruments continues to improve, the tension
between theory and experiment gets bigger, not smaller. There are therefore many good reasons
for continuing to search for a more reliable (and complete) cosmological model.

This could simply be facilitated by the actual measurement of a redshift
drift, which over time would help us identify the values of the unknown parameters
in Equation~(\ref{eq:E}), and thereby eventually helping us identify the theory 
underlying their existence.

It is also possible that the upcoming measurement of redshift drift may 
indicate to us that the expansion of the Universe is not accelerating after all. If the 
measured redshift drift turns out to be non-zero, no matter the value, the proposed 
solutions described in \S~5 above become moot. But if the outcome is unambiguously zero, 
the major problems and inconsistencies plaguing the standard model simply go away. 
There are few instances in science when the anticipated impact of an experiment carries
this much weight.

\References
\item[] Aad G \etal (ATLAS collaboration) 2012 {\it PLB} {\bf 716} 1
\item[] Abreu G \& Visser M 2010 {\it Phys Rev D} {\bf 82} id 044027 10pp
\item[] Adler R J \etal 1995 {\it AJP} {\bf 63} 620
\item[] Alexander T \& Natarajan P 2014 {\it Science} {\bf 345} 1330
\item[] Banados E \etal 2018 {\it Nature} {\bf 553} 473
\item[] Brandenberger R H \& Martin J 2013 {\it CQG} {\bf 30} id 113001 
32pp
\item[] Cadmus R R 1999 {\it AJP} {\bf 67} 665
\item[] Christiansen J L \& Siver A 2012 {\it AJP} {\bf 80} 367
\item[] Cook R J \& Burns M S 2009 {\it AJP} {\bf 77} 59
\item[] Englert F \& Brout R 1964 {\it PRL} {\bf 13} 321
\item[] Guth A H 1981 {\it PRD} {\bf 23} 347
\item[] Harvey A \& Schucking E 2000 {\it AJP} {\bf 68} 723
\item[] Higgs P 1964 {\it PRL} {\bf 13} 508
\item[] Inayoshi K \etal 2016 {\it MNRAS} {\bf 459} 3738
\item[] Kibble TWB 1976 {\it J Math Phys} {\bf 9} 1387
\item[] Kloeckner H R Obreschkow D Martins CJAP \etal 2015
{\it Proceedings Advancing Astrophysics with the Square Kilometre Array} (PoS AASKA14) 027
\item[] Lazanu A Martins CJAP \&  Shellard EPS 2015 {\it PLB} {\bf 747} 426
\item[] Liske J \etal 2008 {\it MNRAS} {\bf 386} 1192
\item[] Liske J \etal 2014 {\it Document ESO 204697 Version 1}
\item[] Liu J \& Melia F 2020 {\it Proc R Soc A} {\bf 476} id 20200364 13pp
\item[] Lovis C \& Pepe F 2007 {\it A\&A} {\bf 468} 1115
\item[] Martin J \& Brandenberger R H 2001 {\it PRD} {\bf 63} id 123501 16pp
\item[] Melia F 2016 {\it MNRAS Lett} {\bf 463} L61
\item[] Melia F 2020 {\it The Cosmic Spacetime} (Oxford: Taylor \& Francis)
\item[] Melia F 2021a {\it EPJ-C} {\bf 81} id 234 12pp
\item[] Melia F 2021b {\it PLB} {\bf 818} id 136362 11pp
\item[] Melia F \& L\'opez-Corredoira M 2018 {\it A\&A} {\bf 610} id A87 5pp
\item[] Nielsen A B \& Visser M 2006 {\it CQG} {\bf 23} 4637
\item[] Perlmutter S \etal 1998 {\it Nature} {\bf 391} 51
\item[] Planck Collaboration \etal 2020 {\it A\&A} {\bf 641} A6 67pp
\item[] Riess A G \etal 1998 {\it AJ} {\bf 116} 1009
\item[] Riess A G \etal 2019 {\it ApJ} {\bf 876} id 85 13pp
\item[] Sandage A 1962 {\it ApJ} {\bf 136} 319
\item[] Schmidt B P \etal 1998 {\it ApJ} {\bf 507} 46
\item[] Sousa L \& Avelino P P 2015 {\it PRD} {\bf 92} id 083520 10pp
\item[] Steinhardt C L \etal 2016 {\it ApJ} {\bf 824} id 21 9pp
\item[] Vilenkin, A. \& Shellard, E.P.S. 1994, {\it Cosmic Strings
and other Topological Defects} (Cambridge: Cambridge University Press)
\item[] Volonteri M \& Rees M J 2005 {\it ApJ} {\bf 633} 624
\item[] Weinberg s 1972 {\it Gravitation and Cosmology: Principles and 
Applications of the General Theory of Relativity} (New York: John Wiley \& Sons)
\item[] Weyl H 1923 {\it Z Phys} {\bf 24} 230
\item[] Zel’dovic Y B Kobzarev I Y \& Okun L B 1975 {\it Sov Phys JETP} {\bf 40} 1
\endrefs

\end{document}